# The right to audit and power asymmetries in algorithm auditing

**Authors**
Aleksandra Urman[1] - University of Zurich, Andreasstrasse 15, 8050 Zurich, Switzerland
Ivan Smirnov - University of Mannheim, 68131 Mannheim, Germany
Jana Lasser - Graz University of Technology (TU Graz), Rechbauerstraße 12, 8010 Graz, Austria

***Abstract***: *In this paper, we engage with and expand on the keynote talk about the "Right to Audit" given by Prof. Christian Sandvig at the IC2S2 2021 through a critical reflection on power asymmetries in the algorithm auditing field. We elaborate on the challenges and asymmetries mentioned by Sandvig - such as those related to legal issues and the disparity between early-career and senior researchers. We also contribute a discussion of the asymmetries that were not covered by Sandvig but that we find critically important: those related to other disparities between researchers, incentive structures related to the access to data from companies, targets of auditing and users and their rights. We also discuss the implications these asymmetries have for algorithm auditing research such as the Western-centrism and the lack of the diversity of perspectives. While we focus on the field of algorithm auditing specifically, we suggest some of the discussed asymmetries affect Computational Social Science more generally and need to be reflected on and addressed.*

**Keywords:** algorithm auditing, computational social science, power asymmetries

In March 2020, a U.S. court ruled that creating fictitious user accounts on employment sites is not a crime [1]. This was the culmination of a four year process that started when Prof. Christian Sandvig from the University of Michigan filed the complaint challenging the Computer Fraud and Abuse Act, the provision that makes it a crime to use a website in violation of its Terms of Service. The story behind this lawsuit was presented by Sandvig in his keynote talk at the International Conference on Computational Social Science (IC2S2) 2021 [2]. In this talk, he spoke about studies based on the algorithm auditing methodology, more specifically *adversarial* algorithm audits - i.e. independent audits that are not commissioned by the creators of the algorithm and might uncover problematic algorithmic behaviour. Sandvig discussed the obstacles scientists face when conducting such studies, predominantly focusing on legal issues. In this paper, we critically engage with and expand on Sandvig's talk. Specifically, we unpack the asymmetries in the capability to take the risk of legal retaliation and access to data for different researchers and how that might lead to biases in which platforms get audited. We also extend the ethical considerations surrounding online auditing studies to platform users and other researchers. As Sandvig's talk was focused on adversarial audits, in this paper when talking about auditing we also mean studies of adversarial nature.

---
[1] Corresponding author. Email: urman@ifi.uzh.ch

## Auditing: definition and origins

Sandvig's talk focuses on algorithm auditing - a form of auditing studies that examines contemporary complex algorithms deployed, for instance, for content recommendations and information filtering in the online realm. One commonly used definition of algorithm audits was given by Mittelstadt [3] - "a process of investigating the functionality and impact of decision-making algorithms" (P. 4994). Mittelstadt also distinguishes between functionality audits, which examine how an algorithm works, and impact audits, which analyse algorithmic outputs and evaluate them for the presence of biases, misrepresentations and other distortions.

Algorithm audit studies have emerged relatively recently, with first corresponding papers published approximately ten years ago. (see [4], [5] for a review). However, in social sciences audit studies of other types have been conducted for decades [6], as Sandvig also notes in his talk. A classic social science audit study is a type of field experiment, i.e. it aims to explore phenomena as they occur naturally rather than in laboratory settings. The goal of the audit study is typically to uncover discrimination, biases, or undesirable effects of a policy, hence, the "audit" name.

A typical example of an audit study is an experiment conducted in the early 1970s in the US [7]. Pairs of observers which were either caucasian or from an ethnic minority visited twenty five apartment houses that were advertised for rent. Observers introduced themselves as couples looking for an apartment. All pairs of observers provided the same backstory and were similar in other aspects. Despite this, less than a half of minority couples were told that apartments are available while for Caucasian pairs, the positive response rate was 80%, indicating ethnic discrimination.

Audit studies do not necessarily involve real people, another classic approach involves so-called correspondence studies, where fictional CVs are sent to potential employers. For example, gendered or ethnic names could be randomly assigned to otherwise identical CVs allowing to compare response rates and identify potential discrimination [8], [9].

Classic social science audit studies, such as those outlined above, have been associated with ethical questions and challenges. In particular, study participants (i.e. landlords, recruiters, etc) do not provide informed consent, they are often lied to, waste their time on fictional applications, and they are often not debriefed. Debriefing with audit studies can cause additional harm to the participants ranging from psychological discomfort (e.g., if a participant rejected a minority candidate and regretted it) to legal consequences.

Despite these ethical concerns, audit studies are generally accepted in social sciences as the risk is considered to be minimal and benefits presumably outweigh the risk [10]. The acceptance of audits is additionally motivated by the fact that there are often no alternative ways to uncover and reliably document discrimination or biases in certain circumstances. Notably, classic audit studies have been deemed legal by the courts [10] despite the associated ethical challenges.

## Algorithm audits

As noted in the beginning, Sandvig's talk and this paper focus on a more novel form of audit studies: algorithm audits. Most commonly algorithm audits are employed to study online platforms such as social media or web search engines to assess the factors that contribute to personalization [4], [11], evaluate potential price discrimination [12], investigate how algorithms affect users' exposure to different types of information [13]–[15], or determine whether certain phenomena or social groups are misrepresented online due to algorithmic information filtering [16], [17]. A comprehensive review of such audits performed on online platforms can be found in [5].

In other societally relevant domains where algorithms are used such as policing, hiring, insurance or banking, algorithm audits have also been performed [18]–[20] though they are less common in these areas. Notably, Sandvig's talk was also centred on online platform audits rather than audits in other domains. We suggest that the lower prevalence of audits in these other domains is not due to their lower importance - it is certainly not the case that discrimination in hiring or policing is less consequential than discrimination on social media or online shops - but rather is the result of the power asymmetries that are relevant in the context of algorithm auditing that we will discuss in the following sections. The biggest challenge in this case - as well as with the online platform audits - comes from the obstacles to accessing data from the private companies that own the platforms and are responsible for creating and using the algorithms the researchers would like to audit. In the next subsection, we briefly summarise the main points and arguments from Sandvig's talk that are relevant in this context.

## Legal restrictions in the context of algorithm auditing and "the right to audit"

According to the experience of Sandvig and his colleagues, private corporations are highly apprehensive of independent audits conducted by scientists. These companies tend to view their platforms and algorithms as private property hosted on their own private servers, and thus posit that scientists have no inherent right to examine this private property and associated data. Sandvig recalls a situation in which scraping data from online platforms was compared by the corporation to entering a physical store and starting digging a hole in the middle of it "for scientific purposes".

This position of the private corporations is often supported by legal actions or threats of legal action against the scientists conducting the audits, particularly in the U.S. context that is the focus of Sandvig's talk. For example, Spotify threatened researchers who revealed its history of pirating content with legal actions [21] for violating its Terms of Use. Such legal actions could lead to grave consequences, as happened in the infamous case of Aaron Swarz who committed suicide facing the possibility of imprisonment after downloading a large number of scientific articles from JSTOR [22]. These are just the examples of the companies threatening researchers with legal action that Sandvig provided in his talk, but there are many others. Some recent examples include Meta going after researchers at New York University [23] or pressuring the Berlin-based NGO AlgorithmWatch to shut down their study of Instagram with threats of legal action [24].

Despite the aforementioned legal issues, in some cases, courts ruled that scraping information that is already publicly available on the internet is legal (e.g., [25]). Moreover, in the case filed by Sandvig (Sandvig v. Barr [1]), the ruling was ultimately in Sandvig's favor, concluding that researchers can not be persecuted for conducting algorithm auditing, in particular through the creation of fake "tester" accounts in order to uncover racial, gender, or other discrimination on online platforms, even if the data collection violates the Terms of Service of online platforms [26]. Nonetheless, in his talk Sandvig argues that the ruling was too narrow in scope. He calls for a broader right to audit which he understands as "independent social research as a necessary minimum requirement for computational systems" [2]. As we understand it, actions taken by researchers to audit computational systems should be not just not forbidden but explicitly allowed, without the possibility of being restricted by platforms (e.g., through their Terms of Service). In the Q&A session after the talk Sandvig explained, for instance, that even despite the court ruling in Sandvig v. Barr the corporations can still threaten researchers with legal action. Even if the court would rule in the researchers' favour in the end, such lawsuits might be too costly for researchers to afford. Companies could also accuse researchers of scientific misconduct as there is no de facto right to audit yet. According to Sandvig, this right should be an uncontroversial issue.

This is because the damage to the platforms from audit studies is lower than damage to the participants of classic audit studies such as landlords while the benefits are higher given the scale of the potential problems. Sandvig posits that the auditing performed by the researchers has negligible impact on the platforms given their vast resources and, in most cases, would be unlikely to be even noticed by the platforms. With the classic audit studies, on the contrary, those audited typically spend a noticeable amount of their work time on the fake requests from the researchers. Thus, Sandvig argues, algorithm audits are less intrusive and have a smaller negative impact on the side that is audited. Additionally, Sandvig suggests that the researchers' right to collect the data from online platforms should be considered uncontroversial because researchers are acting in the public interest.

However, we believe that this issue is more complicated and not as uncontroversial as presented by Sandvig, mainly because of the asymmetry of power between different stakeholders. In his talk, Sandvig touches upon some of those asymmetries but does not mention others that we believe are equally important. In the next section, we unpack these asymmetries.

## Power asymmetries and algorithm auditing

### Asymmetries in the targets of auditing

As was mentioned above, the field of algorithm auditing extends beyond online platforms as algorithms are also used in hiring, banking and other socially important domains. Yet, the audits of algorithms by independent researchers are rare in these domains. We argue that this is largely due to the lack of access to the data by the researchers. While in the case of online platforms researchers often can - at least from a technical though not necessarily legal standpoint - access the data and conduct experiments on algorithmic outputs, in the case of the other domains mentioned above, such access and experimentation is most often not even technically possible for outside researchers. In this case, the audit access lies fully with the private corporations which, as Sandvig notes, are usually apprehensive about sharing their data.

The situation is arguably even more convoluted when it comes to the algorithmic decision-making employed by governments rather than private corporations. Examples of this include policing and immigration. For instance, it is known that the governments of the UK, Canada, New Zealand and some other (Western) countries employ algorithms in their decision-making processes when it comes to visa requests, deportations and/or other immigration-related issues [27]. Researchers have highlighted a number of ways in which such algorithms might be biased based on the examples of algorithmic biases from other domains [27]. Yet, actual algorithm audits that could provide evidence on the absence or presence of such biases and discrimination as a result in immigration-related algorithms - or any other algorithms employed by the governments - do not exist since they are impossible to conduct without access to the data that is handled by government agencies.

To sum up, Sandvig highlights that the access to the data from online platforms is often obstructed by various legal issues. However, in the cases when algorithm-relevant data can not be scraped, the access to it is also technically obstructed, resulting in a major imbalance in terms of the entities whose algorithms get audited at all.

**Asymmetries in prohibiting vs. allowing and incentive structures**

In this section, we will again shift our attention specifically to online platforms and the asymmetries related to the Terms of Service (ToS)-based obstructions in the access to data, as this was the main focus of Sanvig's talk.

As Sandvig notes, platforms usually prohibit scientists from doing audits or other types of studies by adding corresponding statements to their Terms of Service (ToS). These statements can be quite radical, for example: "You shall not use any robot, spider, scraper, data mining tools, data gathering and extraction tools, or other automated means to access our Service for any purpose, except with the prior express permission of [the platform] in writing" – as stated in the ToS of the platform ResearchGate [28]. Sandvig gives even more extreme examples that he and his colleagues encountered, such as the prohibition to "record anything from the platform even with pen and paper" or statements like "no researcher may use this platform". However, he did not provide the names of the platforms that used such radical formulations in their ToS, and we were unable to identify them, even if it is not hard to imagine that they might exist. In other cases, though, ToS are more permissive, especially when it comes to scraping for non-commercial purposes (such as academic research). For instance, in 2021 TikTok's ToS actually did not prohibit scraping of content as long as it was for non-commercial purposes [11]. However, by the end of 2022, the ToS were changed, and now TikTok prohibits the extraction of "any data or content from the Platform using any automated system or software", regardless of the purpose [29].

We suggest that this shift from more permissive to more prohibitive ToS in terms of scraping, similarly to the prohibitive nature of ToS in general, is related to an asymmetry in the incentives and counter-incentives for online platforms in the context of data access. In fact, there is no obvious downside for the platforms to being overprotective about the access to data. Further, including prohibitive statements in ToS is much easier and cheaper than coming up with and enforcing elaborate rules about data access. Thus, permissive ToS are actually counter-incentivized for the platforms. This results in the ToS often prohibiting web scraping even if companies are in fact completely fine with providing data in certain situations (e.g., for non-commercial purposes such as research).

The authors of this paper and their colleagues have had numerous experiences that exemplify this. For example, one of the authors once requested permission to scrape the data from a website. Its owners confirmed that they allow this and were even surprised by the request because they assumed that no permission is required for collecting the data that is already publicly available. However, their ToS published on their website actually prohibited data scraping. In a different case, a representative of a large online platform reached out to the authors of a recently published research study of that platform to talk about it. Though research involved automatically scraping data from the platform - which its ToS prohibit - the representative showed interest in the findings of the research and claimed the platform is not against researchers scraping the data. However, the prohibition could not be excluded from ToS as, according to the representative, the corporation wants to prevent the abuse of ToS by its competitors or other companies who might want to scrape large amounts of data for commercial use. In another case, one of the authors sent an inquiry about the possibility of automatically collecting data from a large online platform. The author received a positive answer from the platform and then asked for a confirmation that could be shown to a journal once the paper is submitted. The initial response was again positive but then the author was told that the legal department of the platform advised against issuing any confirmation.

All the examples above once again demonstrate that the companies often tend to include prohibitive statements in their ToS not necessarily because they are against the use of their data for research, but rather to be on the "safe side" in terms of the legal issues around the access to their data. We suggest that this can be changed only if the asymmetry in the incentives to allow vs. prohibit data scraping for research shifts drastically, and it becomes more attractive for the companies to allow data access rather than forbid it. There are multiple ways this can be achieved, but we suggest the most likely one will be based on a combination of academia-industry collaborations that would make (some) research findings useful enough for the companies to allow data access and new legal regulations that force the companies to provide access to certain types of data to independent auditors and/or academic researchers (similar to the new mandates of the Digital Services Act in the EU [30]).

**Asymmetries in the ability to defend**

The same asymmetry in the incentives to prohibit vs. to allow data collection exists when it comes to the legal departments at universities, Institutional Review Boards (IRBs) or similar ethical committees. They have no incentive to take a risk by allowing something that might turn out to be illegal or unethical. From our experience and the experience of our colleagues, the typical response from the legal department is to prohibit collecting data, and from the IRB is to say that this lies outside their responsibilities and competencies. In such circumstances, researchers often proceed without having a formal approval and therefore explicit protection by their university, and rely on precedents: online data is routinely collected by other researchers after all. Nevertheless, this leaves the individual researcher vulnerable to legal retaliation by the audited corporations. In such a case, a researcher would face an opponent with a large legal department and practically unlimited financial resources. This asymmetry is exacerbated within the research community: while tenured researchers usually have the legal departments of universities to back them, early career researchers (ECRs) oftentimes do not enjoy this privilege. Another concern for researchers is accusations of unethical behaviour. As research is essentially a public activity, success in science heavily relies on reputation and reputation is asymmetrical. No matter how many "ethical" studies a researcher conducted, it is enough to have one study that is considered unethical to ruin a career. For ECRs, their ability to conduct research, support their family and sometimes even their residence permit might depend on getting the next contract. Given these extremely high stakes, ECRs will think twice about engaging in a research project that could potentially lead to backlash from a corporation and endanger their whole research career.

This asymmetry is also noted by Sandvig along with the obvious unfairness of it and negative consequences for the ECRs. Sandvig then also suggests that tenured researchers thus should engage in auditing since such research practices do not pose major risks to their careers. While these are valid points that we agree with, there is an additional negative consequence that Sandvig does not explicitly mention. The current (legal) status-based asymmetry between ECRs and tenured researchers in terms of their ability to conduct audits has negative implications for the field of algorithm auditing itself due to the overrepresentation of white male and more privileged scholars among those tenured [31]. Furthermore, in some disciplines in the U.S. at least, the gender gap among tenured faculty is wider among researchers of foreign origins than among those originally from the country. [32]. We suggest that the power asymmetry between ECRs and tenured scholars along with the lack of diversity among tenured faculty can lead to a bias in the types of auditing studies that are being conducted. Scholars who are not well established yet might shy away from auditing studies because they are too risky. As a result, valuable perspectives that could inform development of the platforms such as they serve all communities may be overlooked. We will return to this point and provide relevant examples in the next section in which we discuss the assumed homogeneity of researchers.

## Going beyond researchers vs. companies

In his talk, Sandvig presents the issue from the researchers vs. corporations perspective. Many of his points rest on the assumption that researchers are a homogenous and benevolent force serving society. In this section, we would like to challenge this assumption as well as discuss its implications for the algorithm auditing field.

**Researchers are not homogeneous**

As noted in the previous section, one aspect of this issue is admittedly mentioned in Sandvig's talk when he discusses the differences in the status and protections enjoyed by tenured researchers compared to ECRs. Another difference among researchers mentioned in Sandvig's talk is that between foreign and non-foreign researchers, since the former face an additional legal risk in the form of potential visa revocation and deportation if accused of illegal activities. However, we suggest that the issue of non-homogeneity of the research community and its implications for the field are broader than what was mentioned by Sandvig. Different groups of researchers face different risks and also benefit differently.

One disparity among researchers that was not mentioned so far is that between scholars from "prestigious" and "not prestigious" institutions. The former are often better connected than the latter, including in terms of connections to the companies that can be helpful to obtain the data necessary for auditing and other studies in the computational social science domain. Further, even when the access to platform data is formalised in a certain form but still gated - in the sense that not everyone can have access to the data automatically but rather only those researchers who have been somehow vetted - the disparities among researchers in terms of prestige and status might play a role. It is difficult to evaluate however whether and to what extent that is the case since the processes behind the decisions to grant or refuse access to the data are usually opaque.

One prominent example is Twitter's Academic API. While the API itself is a great tool that allows researchers to conduct all kinds of research using Twitter data – including research into the workings of Twitter itself – access to the API requires an application process. The outcome of the application is rather unpredictable and no justification is given for negative outcomes. Moreover, once access is denied, no re-application or challenge of the outcome is possible. Our and our colleagues personal experiences highlight the opacity in the process: some of us were denied access to the API from the start, others were first approved, then at some point blocked from the access, and some years later reinstated without a reason, all of this without any explanations about the decisions from the side of Twitter. We have helped numerous students to apply for API access, sometimes with virtually identical applications, some of which got denied while others were approved immediately - an experience which could be framed as an auditing study itself.

Another example is the Social Science One initiative that aims to foster collaboration between researchers and companies. Specifically, it enables access to Facebook data for (selected) researchers. Data access is given or refused based on a process similar to obtaining research funding: researchers submit project proposals, and then are granted or refused access to data based on the evaluation of the proposals according to "academic merit and feasibility; research ethics; likelihood of knowledge resulting from the project advancing social good; qualifications of the proposed team" [33]. It is however unclear how exactly the proposals are evaluated, which is problematic as some of the criteria - such as the likelihood of the project to contribute to social good - are rather subjective. Others, such as the evaluation of researchers' qualifications, can be influenced by implicit and explicit biases (e.g., related to the applicant's gender), as studies based on academic funding award procedures show [34]. In the case of Social Science One there is an apparent gender disparity among those awarded access to the data. In the first cohort of data access awardees - the only cohort on which we found a publicly available list of awardees - only 2 out of 13 Principle Investigators appear to be women[2], one of them leading a project together with a male PI [35]. Due to the opacity of the evaluation process and the unavailability of the information on the application success rate, we can only point out this gender disparity as a fact but not infer its actual cause. While evaluation processes of public research funding agencies have similar biases [34], information necessary to scrutinise the processes is more widely available or can be acquired via freedom of information requests or similar laws.

Restrictive processes to regulate access to platform data by academics are often justified by protecting user privacy or preventing data misuse. However, to date there is no evidence that such screening procedures actually prevent any wrong-doing. It seems like what matters most for being granted access to data is the nationality or country of residence of the researchers, the prestige of their institution and their connections to the corporation providing data access. As a result, initiatives that are often considered as equalising science (e.g. Twitter API allows anyone to get access to a lot of data and do research), can in fact exacerbate inequalities, which in turn affects the field of algorithm auditing itself.

**Implications of the lack of diversity among researchers**

We suggest that the aforementioned differences among researchers - i.e., those related to the ease of data access and risks associated with engaging in risky algorithm audits - in terms of the scholars' gender, status, nationality, ethnicity or prestige of their institutions, lead to the striking imbalances in the field of algorithm auditing with regard to what gets audited. Specifically, these differences contribute to the field being highly Western- and English-centric as those for whom audits are less risky or easier to implement most often are based at prestigious Western institutions, more likely to be white and male, and less likely to have a migration background. A recent literature review of (online platforms focused) algorithm audits [5] and a search for "algorithm audit studies" in a scholarly database indicate that the absolute majority of the studies conducted to date focus on 1) Western platforms and 2) content published in English language. There are notable exceptions to this: several studies, authored by academics based in German-speaking countries, have examined content in German [36]–[38]. A few others either adopted a comparative approach analysing content across multiple languages [13] or audited non-Western platforms and/or focused on content in languages such as Russian or Chinese [39]–[45]. However, these studies are exceptions to the general "rule" of the Western- and English-centric studies in the algorithm auditing field.

Since most large online platforms are, in fact, Western and predominantly, U.S.-based, one could argue that Western-centrism of algorithm audits is not a major issue. In the end, it is plausible to assume - though we do not know this for a fact - that there are no separate algorithms deployed by companies in different jurisdictions and local markets. If this assumption is correct, the analyses of the factors that contribute to personalization on Google or TikTok [11], [12] are valid not only for a certain national context, thus Western-centrism should not be much of an issue for such functionality audits. Nonetheless, we suggest it is problematic that, with the exception of TikTok, Baidu or Yandex [39]–[45], no other platforms that were founded outside of the U.S., to the best of our knowledge, have been subjects of algorithm audits so far. Since many of such platforms are founded and based in authoritarian states such as Russia or China and some are known to cooperate with the governments to censor or downgrade political content, auditing them can bring particularly valuable insights about potential algorithmic biases and authoritarian manipulation.

---

[2] We acknowledge that gender is a complex construct and as we relied on visual cues and publicly available CVs only when determining the gender of the awardees, we can not be certain about the gender identities of the PIs, hence we use the term "appear to be" women instead of "are" women.

In the context of algorithm impact audits, analyses of non-Western platforms and non-English content are arguably even more overdue. Even if the algorithms deployed by the companies across the globe function in the same way in different regions, the pool of content available for an algorithm to select and (de)prioritise from differs depending on the language. Thus, even an algorithm that works in the same way will yield qualitatively different sets of results depending on the language. Further, it is known that the companies tend to invest more resources into content moderation and annotation (that is used as training data for machine learning algorithms) for English language content and, to a lesser degree, other popular and/or Western languages [46]. Taken together, the differences in the pool of content and in information quality assurance-related efforts depending on the language, result in information inequalities between users depending on their location and/or language. This is documented by the few existing comparative algorithm audits (e.g., [13], [39]), with the inequalities sometimes being potentially life-threatening for the users. For instance, Scherr and colleagues found that there are major linguistic and location-based differences in how frequently information about suicide helplines is presented to Google's users when submitting suicide-related queries, to the point where in some countries such crisis-prevention functionality seems to be not implemented at all [13]. Such documented information disparities highlight the urgency of conducting more comparative and/or non-English content-focused algorithm audits and giving higher prominence to the fact that the findings of most algorithm impact audits are context-specific, with their findings non-generalizable to other national and linguistic settings.

**Researchers are not necessarily a benevolent force serving larger society**

Coming back to the assumed homogeneity among researchers and Sandvig's talk, we suggest that one aspect that is ignored by Sandvig is an obvious conflict of interest that arises when researchers ask for the possibility to conduct experiments or to access the data. It is important to remember that the main beneficiaries of research are researchers themselves. They don't do research as volunteers or publish their findings anonymously. On the contrary, they are paid for their research and their income and status depend on the visibility of their research. Getting access to unique large scale data sets increases the chances of publication in prestigious journals, promotion, etc. Moreover, researchers are not only interested in accessing the data but they are also interested in restricting data access for other researchers to prevent being "scooped". This becomes especially problematic when researchers that have a stake in research that is conducted using a corporation's data are part of an advisory board or other decision mechanism that decides who gets access to the data. While these self-interested motivations of researchers are not necessarily in contradiction with the interests of society as a whole, we suggest it is important to keep in mind that serving the larger society is often not the main motivation of researchers when embarking on a new project.

## Users and their rights

The omission of a key stakeholder in the algorithm audit debate becomes all the more important when considering that researchers may not be primarily motivated by serving society. Specifically, in Sandvig's talk users of the audited platforms were not discussed. This could be due to the perspective that users in the scholars vs. corporations debate are implicitly assumed to be represented by researchers. This would be in line with the assumption that researchers are acting in the interests of the broader public by default. However, because of the aforementioned potential conflict of interest, we argue researchers in fact do not represent users and their interests do not necessarily align with the interests of the users.

We illustrate this misalignment of interests between researchers and users with the example of sock-puppet auditing studies: In algorithm auditing studies the creation of fake accounts on an online platform is a component of a commonly utilised auditing method, sometimes referred to as "sock-puppet auditing" [47] or "agent-based testing" [48], [49]. However, the creation of fake accounts can have potentially negative effects not only on the platforms - something Sandvig discusses - but also on the users. Sandvig argues that online audits should be uncontroversial because offline audits are uncontroversial and well-established in social sciences as well as less obtrusive towards those audited - in this case, online platforms. This however ignores asymmetries that are introduced by the online environment. Compared to the offline settings - and classic audits conducted by the social scientists - in the online settings, the researchers leave (more) traces that can get in the way of user experiences. In fact, online researchers might leave permanent traces such as fake posts on platforms or buggy commits to software [50]. Researchers therefore not only use a companies' resources but also permanently alter the environment that users experience. From an ethical perspective, the possible harm of such alterations needs to be balanced with the possible benefit and put into perspective with alterations that other actors on such platforms regularly introduce. When researchers pay to distribute an ad over the ad distribution service of a social media platform to test a platform's audience targeting algorithms, it will likely be one of many ads that a user sees during one usage session of the platform. Such an alteration therefore does not introduce an experience that is drastically different from the everyday experience a user has and might be considered as not very harmful. On the other hand, if a researcher introduces offensive content into a self-help forum used by a vulnerable group of people to audit the platform's content moderation policies, users might be confronted with content that is drastically different from what they are used to seeing and might react very negatively to it.

Along the same lines, researchers also need to be careful to not "poison the well" for other researchers that want to conduct (auditing) studies on the same platform: if researcher's auditing behaviour is too obvious or disturbs normal operations too much, other research projects might be endangered because users start recognizing researcher's behaviours and alter their own behaviour as a reaction.

Last but not least, so far users' perspectives have been barely integrated into the design of algorithm audits. Yet recent research suggests that users are able to identify potential algorithmic harms, albeit which harms exactly they identify largely depends on their experiences and demographics [51]. This once again highlights the importance of the diversity of perspectives for the algorithm auditing field. DeVos and colleagues suggest that user perspectives can be integrated into algorithm auditing in the future and advocate for user-driven algorithm audits - while acknowledging potential limitations and harms [51]. We share their perspective and suggest that user-centric audits could help alleviate the current imbalances in the field that we described above, particularly with regard to what gets audited, and thus constitute a fruitful direction for future audits.

## Conclusion

In this paper, we have aimed to critically engage with Prof. Christian Sandvig's talk on the "right to audit" at IC2S2 2021 through the discussion of power asymmetries that are important in the algorithm auditing field. We have elaborated on some asymmetries and challenges mentioned by Sandvig - such as those related to legal issues and the disparity between early-career and senior researchers. At the same time, we have discussed power asymmetries that were not mentioned in the original talk but that we find critically important: those related to other disparities between researchers, incentive structures related to the access to data from companies, targets of auditing and users and their rights.

We have highlighted that while in recent years there have been some legal decisions that were in favour of researchers in the context of adversarial auditing such as the ruling in Sandvig v. Barr, the field of algorithm auditing remains somewhat of a grey zone in terms of law which obstructs the ability of the researchers to conduct auditing studies. One major issue that scientists face with regard to this is the risk of legal retaliation or accusations of scientific misconduct from corporations. As Sandvig also highlights in his talk, this issue can discourage researchers, especially those in more precarious positions such as ECRs or foreign citizens, from conducting audits. This and other power asymmetries we discussed above that lead to inequalities in access to data and ability to perform audits have negative consequences for the field as a whole resulting in it being western-centric and lacking diversity of perspectives. Another aspect that needs to be considered in connection to auditing is the ethical implications of conducting (online) audits. While in Sandvig's talk these have been considered for corporations, the needs and concerns of users and other researchers working on the platforms that researchers audit have not been adequately addressed. It is essential to consider the distribution of power among all stakeholders in order to ensure a truly fair and unbiased field of algorithm auditing.

Finally, we would like to note that even though we discussed all the aforementioned power asymmetries in the context of algorithm auditing research, many of them in fact affect Computational Social Science (CSS) more generally with the field as a whole potentially lacking the diversity of perspectives and objects of analysis. We suggest that not only researchers interested in algorithm auditing but the CSS field as a whole needs to critically reflect on the implications of existing disparities and look for the ways to address them.

## Declarations

**Availability of data and materials**
Data sharing is not applicable to this article as no datasets were generated or analysed during the current study.
**Competing interests**
The authors declare that they have no competing interests.
**Funding**
*J.L. acknowledges funding from the Marie Skłodowska-Curie grant No. 101026507.*
**Authors' contributions**
AU, IS and JL came up with the idea for the manuscript together; IS outlined the initial structure of the manuscript and came up with a set of notes that served as a basis of the manuscript, AU and JL contributed additional notes and relevant literature; AU wrote the first full draft of the manuscript and finalised the manuscript based on the comments, edits and additions from IS and JL to the original draft version.